\documentclass[twocolumn,english,aps,prb,showpacs,supperscripaddress,floatfix,final]{revtex4-1}
\usepackage[T1]{fontenc}
\usepackage[latin9]{inputenc}
\usepackage{color}
\usepackage{float}
\usepackage{amsbsy}
\usepackage{graphicx}
\usepackage{esint}

\makeatletter

\@ifundefined{textcolor}{}
{%
 \definecolor{BLACK}{gray}{0}
 \definecolor{WHITE}{gray}{1}
 \definecolor{RED}{rgb}{1,0,0}
 \definecolor{GREEN}{rgb}{0,1,0}
 \definecolor{BLUE}{rgb}{0,0,1}
 \definecolor{CYAN}{cmyk}{1,0,0,0}
 \definecolor{MAGENTA}{cmyk}{0,1,0,0}
 \definecolor{YELLOW}{cmyk}{0,0,1,0}
}

\makeatother

\usepackage{babel}
\begin{document}

\title{Landau-Ginzburg Theory for Anderson localization}

\author{Samiyeh Mahmoudian}

\affiliation{Department of Physics and National High Magnetic Field Laboratory,
Florida State University, Tallahassee, Florida 32306, USA.}

\author{Vladimir Dobrosavljevi\'{c}}

\affiliation{Department of Physics and National High Magnetic Field Laboratory,
Florida State University, Tallahassee, Florida 32306, USA.}
\begin{abstract}
The Typical Medium Theory provides conceptually the simplest order
parameter description of Anderson localization by self-consistently
calculating the geometrically-averaged (typical) local density of
states (LDOS). Here we show how spatial correlations can also be captured
within such a self-consistent theory, by utilizing the standard Landau
method of allowing for (slow) spatial fluctuations of the order parameter,
and performing an appropriate gradient expansion. Our theoretical
results provide insight into recent STM experiments, which were used
to visualize the spatially fluctuating electronic wave functions near
the metal insulator transition in $Ga_{1-x}Mn_{x}As.$ We show that,
within our theory, all features of the experiment can be accounted
for by considering a model of disorder renormalized by long-range
Coulomb interactions. This includes the pseudogap formation, the $C(R)\sim\frac{1}{R}$
form of the LDOS autocorrelations function, and the $\zeta\sim\frac{1}{E}$
energy dependence of the correlation length at criticality.
\end{abstract}
\maketitle
\textit{Introduction-} Until very recently, many experiments on metal-insulator
transition concentrate on describing the bulk behavior averaged over
the entire sample instead of local features on the atomic scale. That
is why, the most theoretical approaches so far focused on global quantities,
such as the sample-averaged transport and thermodynamic behaviors,
which are insufficient to properly interpret the new experimental
data. However, a new generation of STM experiments are suggesting
a deep understanding of the features of the local electronic wave
functions in the vicinity of the metal-insulator transitions \cite{richardella2010visualizing}.
Visualizing the puzzling features of the experiment on the nano-scale
encouraged us to seek for the theoretical framework to describe the
nature of metal insulator transition. For example, one of the spectacular
unexpected result on $Ga_{1-x}Mn_{x}As$ was the emergence of pseudo-gap
at Fermi energy, at the critical concentration, that defy the commonly
accepted picture of how disorder affects the electronic states in
the critical region \cite{anderson1958absence}. This feature of local
quantum wave functions calls for a completely new look at the mechanisms
for localization in the presence of interaction. The challenging question
here is what type of theoretical framework can be sufficient and effective
in order to describe the experimental features in relevant regimes.
As it can be discovered by the experimental data, the size of the
correlation length is comparable to microscopic cut off (the size
of individual manganese atoms). Moreover, the range of energies, which
can be explored by STM measurement, is not much lower than the Fermi
energy. Thus, the Typical Medium Theory such as any other mean field
approaches is applicable to understand the critical characteristic
of this system \cite{mahmoudian2014quantum}. This theoretical approach
can be regarded as an appropriate order parameter theory for Anderson
localization and has the potential to directly combine with already
well accepted Dynamical Mean-Field Theories for strong correlations
\cite{dobrosavljevic1997mean,PhysRevLett.102.156402,aguiaretal1}.
This local order parameter theory has already met with considerable
success in interpreting several experimental observation such as opening
the pseudo-gap \cite{mahmoudian2014quantum}, but so far were not
capable to capture spatial correlation effects. To overcome this obstacle,
we make a key conceptual observation guided by experience in standard
critical phenomena. This standard method has been utilized for describing
the variation of order parameter near the critical point in the spirit
of many mean field theories. For example, it provides a systematic
way to understand how magnetization varies in space in response to
an external field in weiss mean field theory \cite{goldenfeld1992lectures}.
In this letter, we construct Landau-Ginzburg theory of TMT for the
Anderson localization by obtaining corresponding mean-field description
of spatial correlation of the order parameter near the critical point. 

\begin{figure}[h]
\includegraphics[width=3in]{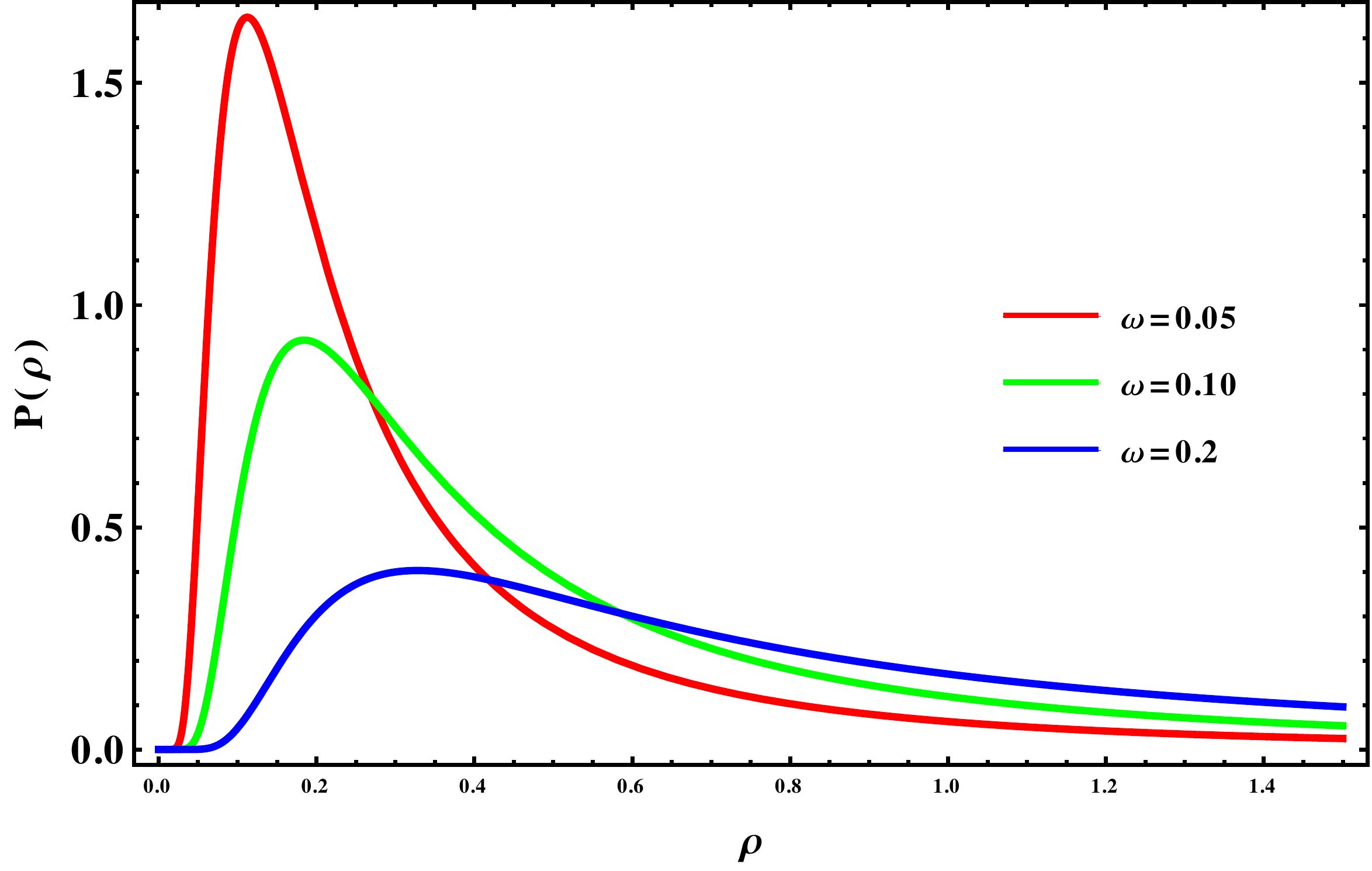}

\protect\caption{Normalized distribution of typical density of states (TDOS) for $\omega=0.05,0.1,0.2.$
As the transition is approached, the most probable value of $P(\rho)$
is reduced . }
\end{figure}

\textit{Model}- Here, we study a single band tight binding model of
non-interacting electrons with random site energies $\epsilon_{i}$
with a given distribution $P(\epsilon_{i})$\textcolor{black}{, which
the Hamiltonian of this system can be written as:}\textcolor{blue}{{} }

\begin{equation}
H=\sum_{\left\langle ij\right\rangle ,\sigma}t_{ij}c_{i\sigma}^{\dagger}c_{j\sigma}+\sum_{i,\sigma}\varepsilon_{i}c_{i\sigma}^{\dagger}c_{i\sigma}.
\end{equation}
Here, $c_{i\sigma}^{\dagger}$ and $c_{i\sigma}$ are the electron
creation and annihilation operators, and $t_{ij}$ are the inter-site
hopping elements. As it has been discussed in our previous work the
\emph{effective} disorder potential seen by quasi-particles can be
significantly modified in presence of \emph{long-range }Coulomb interactions
. \textcolor{black}{This fact leads }to the formation of the soft
\emph{``Coulomb pseudo gap'' }at the Fermi energy and allows us
to consider a model distribution of random site energies which assumes
a pseudo-gap form expected from the Efros and Shklovskii (ES) picture
\cite{efros1975coulomb,efros1976coulomb,efros1985electron}. It is
well known that the statistic of local density of states LDOS reflects
the degree of localization of quantum wave functions \cite{re:Janssen98},
and its most probable, typical value, (TDOS) is represented by the
geometric average of LDOS. Within local TMT approach so far has been
proven that TDOS can be introduced as an appropriate uniform order
parameter. The measurement of this quantity provides the qualitative
feature of density of states \cite{mahmoudian2014quantum}which is
in agreement with both experiment \cite{richardella2010visualizing}
and recent large-scale exact diagonalization results \cite{amini2013multifractality}.
As it has been shown in Fig. (1), near the metal insulator transition
the distributions of (TDOS) begin to cross over from Gaussian to log
normal distribution. Our result, which is consistent with experimental
visualization of local quantum wave functions, can be another reason
for justification of our simple theory. The missing key point in our
previous work was ignoring the spatial correlation effects which has
to be treated beyond the local TMT approach. The challenging question
is how we can constrict the nonlocal order parameter for Anderson
localization starting with the already exciting TMT we earlier proposed
and tested for different systems\cite{0295-5075-62-1-076,mahmoudian2014quantum,PhysRevB.90.094208,PhysRevB.89.081107}.
This is standardly achieved by introducing a weak, spatially varying
external field, and performing an appropriate gradient expansion for
the order parameter, which is the subject of this work and can be
discussed as follows.

\begin{figure}[H]
\includegraphics[width=3in]{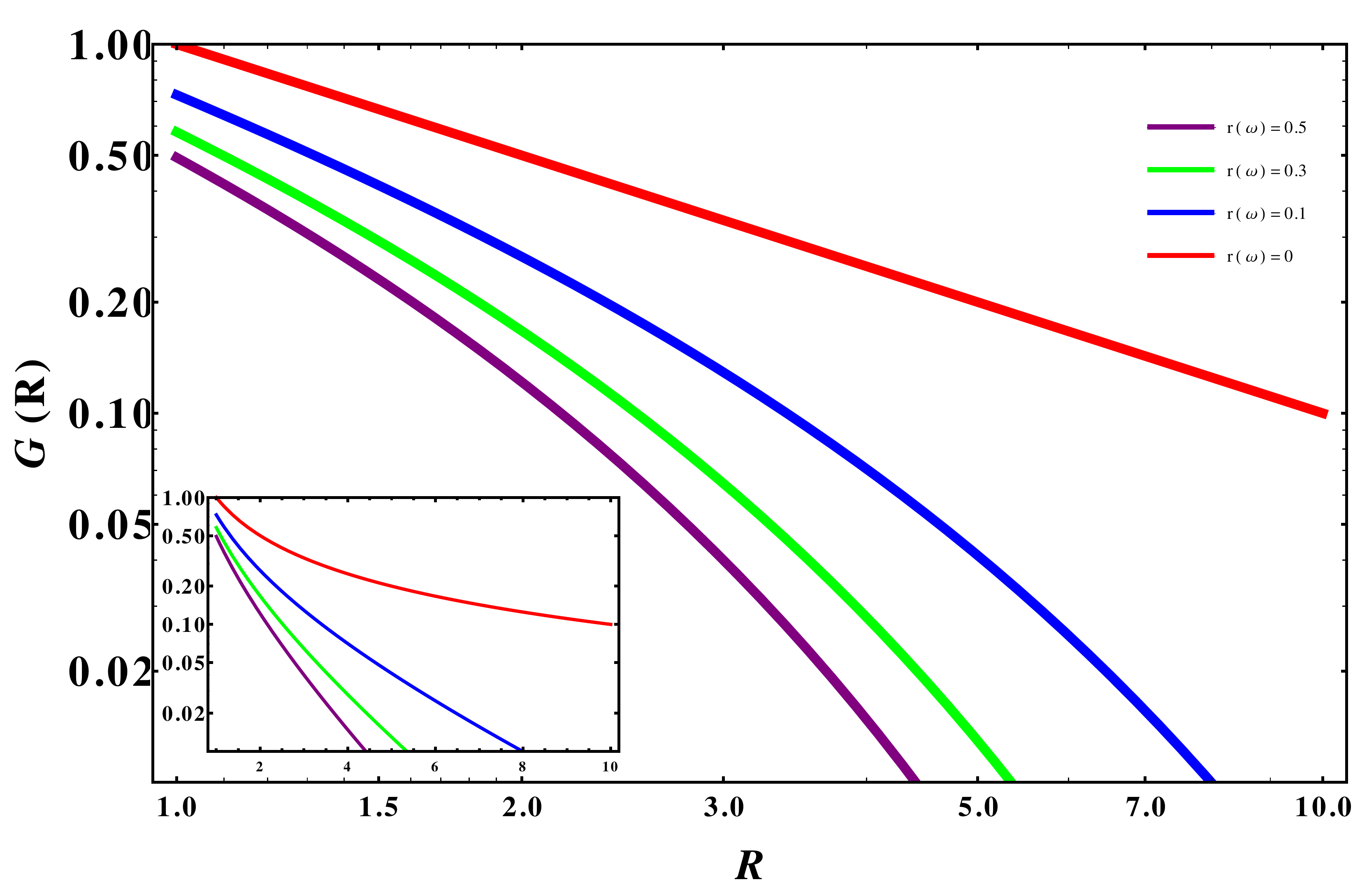}

\protect\caption{The long distance behavior of the Green's function at Fermi level
and also the nearby energies, $r(\omega)=0.1,$$r(\omega)=0.3,$$r(\omega)=0.5$.
The inset shows the same behavior in semi-logarithmic scale.}
\end{figure}

\textit{Derivation of Landau Ginzburg theory- }The general strategy
in formulating an parameter theory follows the ``cavity'' method
typically used in Dynamical Mean Field Theory approaches \cite{georges1996dynamical}.
Here, the dynamics of an electron on a given site can be obtained
by integrating out all the other sites, and replacing its environment
by an appropriately averaged `` effective medium'' $\Delta_{i}(\omega)$,
which is characterized by a self energy $\boldsymbol{\hat{\Sigma}}(\omega).$
The self energy corresponds to this Landau type of TMT is local but
it is not the same for all lattice sites. This property of self-energy
is different from the local TMT, which has been discussed extensively
so far \cite{0295-5075-62-1-076,mahmoudian2014quantum}.

\begin{equation}
G_{ii}^{EM}(\omega)=\left[\omega\hat{1}-\hat{H_{0}}-\boldsymbol{\hat{\Sigma}}(\omega)\right]_{ii}^{-1}.\label{eq:selfenergy}
\end{equation}

Here, $\hat{H_{0}}$ is a bare lattice Hamiltonian matrix and the
``effective medium Green's function'' $G_{ii}^{EM}(\omega)$ is
given by Eq. (\ref{eq:selfenergy}), which it is determined by the
diagonal elements of lattice Green's function. Therefore, in contrast
to local TMT, the effective medium varies slowly in space. Assuming
that the space dependent self-energy is provided at each site $\mathbf{\Sigma_{i}}(\omega)$,
one can calculate the ``nonuniform cavity'' by Eq.(\ref{eq:nonuniformcavity})
as 

\begin{equation}
\Delta_{i}(\omega)=\omega+i\eta-\Sigma_{i}(\omega)-\frac{1}{G_{ii}^{EM}(\omega)}.\label{eq:nonuniformcavity}
\end{equation}

Given the cavity field, the corresponding local density of states
$\rho_{i}(\omega,\epsilon_{i})$ is given by the imaginary part of
the local Green's function: 
\begin{equation}
\rho_{i}(\omega,\epsilon_{i})=-\frac{1}{\pi}Im[\omega+i\eta-\epsilon_{i}-\Delta_{i}(\omega)]^{-1}.\label{eq:5}
\end{equation}

Within Landau Ginzberg describtion of TMT, the\emph{ typical value}
of the local density of states $\rho_{i}^{typ}(\omega)$ still can
be selected as an appropriate order parameter such as standard TMT
approach. However, the modified order parameter has smooth spatial
variations and it is well-represented by the geometric average of
local density of states as 

\begin{equation}
\rho_{i}^{typ}(\omega)=\exp\left[\int d\epsilon_{i}P(\epsilon_{i})\ln\rho_{i}(\omega,\epsilon_{i})\right].\label{eq:4}
\end{equation}

Here, in order to obey causality, the corresponding\emph{ ``typical
local'' }Green's function, is defined \cite{pastor2001tmt,dobrosavljevic2010typical}
by performing the Hilbert transform

\begin{equation}
G_{i}^{typ}(\omega)=\int_{-\infty}^{\infty}d\omega^{\prime}\frac{\rho_{i}^{typ}(\omega^{\prime})}{\omega+i\eta-\omega^{\prime}},\label{eq:Gtyp}
\end{equation}

The selfconsistent loop can be closed \textcolor{black}{by setting
the diagonal elements of the lattice Green function of the effective
medium, $G_{ii}^{EM}(\omega)$ to be equal to the local typical Green's
function} \cite{pastor2001tmt,dobrosavljevic2010typical}\textcolor{black}{.This
yields to the relation}

\begin{equation}
G_{ii}^{EM}(\omega)\equiv G_{i}^{typ}(\omega),\label{eq:latticeGreen's function}
\end{equation}

Therefore, the self-energy can be obtained by the selfconsistency
condition Eq.(\ref{eq:latticeGreen's function}) and Eq.(\ref{eq:selfenergy})
in an iterative scheme as

\[
\Sigma_{i}(\omega)=\omega+i\eta-\Delta_{i}(\omega)-\frac{1}{G_{i}^{typ}(\omega)}.
\]

This method provides us an opportunity to include smooth spatial variation
of order parameter and calculate it in a selfconsistent way. This
Landau description of TMT can be obtained for any general form of
density of states with different random distribution to explain the
second order phase transition. This set of TMT self-consistent equations
can be solved numerically given the bare lattice Hamiltonian $\hat{H_{0}}$
and the form of random potential. In this letter, we are interested
in inventing the analytical approach to describe the critical feature
of the system such as the behavior of correlation function and correlation
length. The full analytical solution of TMT equations for uniform
order parameter has been provided for any general model \cite{mahmoudian2014quantum}.
Although we have found the logarithmic correction to leading critical
behavior of uniform order parameter at the critical point, It has
been proven that this corrections are not important close to the metal
insulator transition\cite{mahmoudian2014quantum}. Knowing this fact,
we are allowed to ignore these mild corrections to develop the non
uniform description of TMT theory. The fact that the order parameter
varies slowly in space as the transition is approached, allows us
to perform the gradient expansion for order parameter $\rho_{i}^{typ}(\omega)$.
Here, we obtain the Landau Ginzberg equation which is given by Eq.(\ref{eq:LG equation})
and provides us the spacial dependance of order parameter in continuum
limit. For the sake of simplicity in notation, we set $\rho_{k}^{typ}(\omega)\equiv\rho.$

\begin{equation}
-\nabla^{2}\rho+r(\omega)\rho+u_{1}(\omega)\rho^{2}+u_{2}(\omega)\rho^{3}+...=0.\label{eq:LG equation}
\end{equation}

Here, 
\begin{equation}
r(\omega)=(W-W_{c})^{\alpha}+\beta\omega^{2},\label{eq:r(omega)}
\end{equation}
 where the critical exponent $\alpha$ is characterized by the form
of renormalized disorder potential. It has been computed for two distinct
classes of random distribution as $\alpha=1$ (\textcolor{black}{generic
model, where $P(\omega)$}\textcolor{red}{{} }\textcolor{black}{is finite
for all $\omega$}) and $\alpha=\frac{1}{2}$ (\textcolor{black}{pseudogap
model, where $P(\omega)$}\textcolor{red}{{} }\textcolor{black}{vanishes
at zero frequency}). The value of $\beta$ is finite at the transition.
As it has been discussed in our previous works \cite{mahmoudian2014quantum,pastor2001tmt},
\begin{equation}
u_{1}(\omega)\propto(P(\omega)+\gamma\omega^{2}),\label{eq:u1(omega)}
\end{equation}
 where $\gamma$ is finite and can be calculated close to the transition
for a given desired distribution. The coefficient $u_{2}(\omega)$
always remains finite for any model of distribution. As an example,
$u_{1}(\omega)$ is finite for the generic model at the transition
$(P(0)\neq0)$. However, for the pseudogap model at the critical point
$(W=W_{c}),$ it is zero at the center of the band $(\omega=0)$ while
it remains finite at nearby energies. These Landau coefficients in
general are complicated which their form depends on both model of
density and disorder distribution. The explicit form of these has
been provided for a general model of density of states with two classes
of random distribution (generic and pseudogap) \cite{mahmoudian2014quantum}.
Interestingly, both model provide the same long distance behavior
for autocorrelation function. Within this framework, we predict that
this function falls off exponentially in both insulating side and
metallic side while it shows the power law behavior at the critical
point $(W=W_{c})$. 

For the purpose of this work, we limit out attention on pseudogap
model of random site energies which is relevant to experimental data.
Given this class of random distribution and predict all feature of
experiment by solving the Eq.(\ref{eq:LG equation}) exactly at the
critical point and also close to the metal insulator transition.

\begin{figure}[H]
\centering{}\includegraphics[width=2in]{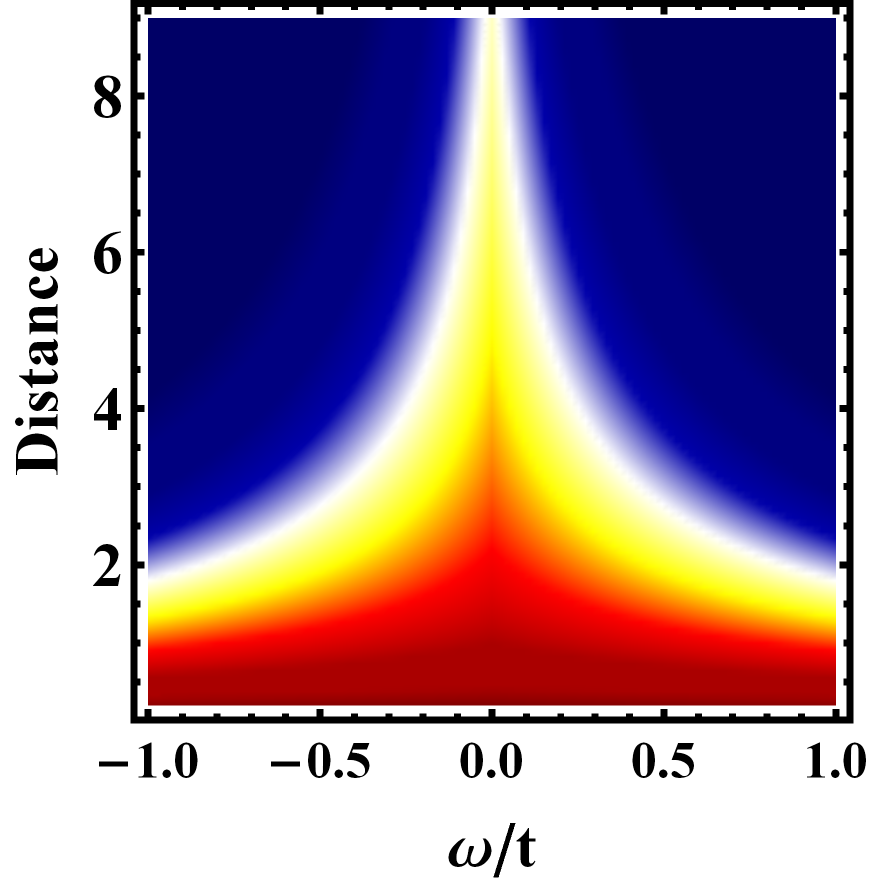}\protect\caption{The behavior of correlation length shows the divergence as $\zeta\sim\frac{1}{\omega}$
at critical point $W=W_{c}.$}
\end{figure}

\textit{Solving the differential equation for pseudogap model}.-We
provide the long distance behavior of autocorrelation function by
providing boundary condition and solving the corresponding differential
equation. We propose that the boundary condition is provided by having
two leads at both metallic and insulating side. This condition is
equivalent with exposing the order parameter to external field as
the standard Landau procedure. The solution can be obtained for any
model of random distribution in different dimension. However, we focus
on the model which is relevant to experiment and can describe the
pseudo-gap formation of order parameter as the metal-insulator transition
is approached for this model. Assuming that the system is uniform
in three dimension, the Eq.(\ref{eq:LG equation}) can be solved for
this model. Within LG description of TMT, we found that at finite
$r(\omega)$ the asymptotic behavior of local density of states autocorrelation
function, which is correspond to Green's function, shows exponential
decays as

\begin{equation}
G(R)=\frac{\exp\{\sqrt{-r(\omega)}R\}}{R}.\label{eq:meal-insulatorG}
\end{equation}
However, at critical disorder $W=W_{c}$ and $\omega=0$, the behavior
of the Green's function is power law and is written as

\begin{equation}
G(R)=\frac{1}{R^{d-2}}=\frac{1}{R}.\label{eq:critical G}
\end{equation}

The behavior of the autocorrelation function has been shown in Fig.
2. The authors also computed the angle-averaged autocorrelation function
between two points separated by $r$. This function decays in space
as $\chi(E_{f},r)=r^{-\eta}$ following power-law with $\eta=1.2\pm0.3$
for the sample with 1.5\% of Mn concentration. while, it falls off
exponentially at all nearby energies \cite{richardella2010visualizing}.
Within LG description of TMT, we found the divergence of correlation
length at critical point $W=W_{c}$ as the signature of metal-insulator
transition. This characteristic can be obtained by the expression
of $r(\omega)$ which is function of distance from transition and
has been shown in Fig. (3). This picture also it in agreement with
experimental measurement of correlation length at critical concentration.
The size of correlation length is in the order of two or three individual
Mn atoms $(20A^{0})$ which justifies the implementing of the TMT
as a sufficient theory to capture nature of the phase transition. 

\textit{Conclusion and outlook.-} In this letter, we combine the simplest
theoretical framework, describing the disorder-driven metal insulator
transition, with standard Landau procedure of order parameter. Within
our invented theory, we predicted the several puzzling features of
experiment on $Ga_{1-x}Mn_{x}As.$ We found the pseudogap formation
of local density of states at the Fermi energy and vicinity of transition.
We discovered that the LDOS autocorrelation functions show power-law
fashion at Fermi energy in large distances. In addition, we verified
the nature of second order phase transition by examine the corresponding
correlation length at criticality. We are suggesting that one should
measure the frequency dependence of entire distribution of LDOS as
it can be computed by our theory. We claim that the entire family
of curves which can be obtained based on existing experimental data,
will collapse by taking our simple theory. It would be also thoughtful
to collect experimental data on insulating side, where the quantum
effects do not play very important rules, in order to be compared
with our theoretical predictions.

The authors thanks Ali Yazdani and Shao Tang for useful discussion.
This work was supported by the NSF grants DMR-1005751, DMR-1410132
and PHYS-1066293, by the National High Magnetic Field Laboratory.

\bibliographystyle{apsrev}
\bibliography{paper,LGpaper}

\begin{thebibliography}{18}
\expandafter\ifx\csname natexlab\endcsname\relax\def\natexlab#1{#1}\fi
\expandafter\ifx\csname bibnamefont\endcsname\relax
  \def\bibnamefont#1{#1}\fi
\expandafter\ifx\csname bibfnamefont\endcsname\relax
  \def\bibfnamefont#1{#1}\fi
\expandafter\ifx\csname citenamefont\endcsname\relax
  \def\citenamefont#1{#1}\fi
\expandafter\ifx\csname url\endcsname\relax
  \def\url#1{\texttt{#1}}\fi
\expandafter\ifx\csname urlprefix\endcsname\relax\def\urlprefix{URL }\fi
\providecommand{\bibinfo}[2]{#2}
\providecommand{\eprint}[2][]{\url{#2}}

\bibitem[{\citenamefont{Richardella et~al.}(2010)\citenamefont{Richardella,
  Roushan, Mack, Zhou, Huse, Awschalom, and
  Yazdani}}]{richardella2010visualizing}
\bibinfo{author}{\bibfnamefont{A.}~\bibnamefont{Richardella}},
  \bibinfo{author}{\bibfnamefont{P.}~\bibnamefont{Roushan}},
  \bibinfo{author}{\bibfnamefont{S.}~\bibnamefont{Mack}},
  \bibinfo{author}{\bibfnamefont{B.}~\bibnamefont{Zhou}},
  \bibinfo{author}{\bibfnamefont{D.}~\bibnamefont{Huse}},
  \bibinfo{author}{\bibfnamefont{D.}~\bibnamefont{Awschalom}},
  \bibnamefont{and} \bibinfo{author}{\bibfnamefont{A.}~\bibnamefont{Yazdani}},
  \bibinfo{journal}{Science} \textbf{\bibinfo{volume}{327}},
  \bibinfo{pages}{665} (\bibinfo{year}{2010}).

\bibitem[{\citenamefont{Anderson}(1958)}]{anderson1958absence}
\bibinfo{author}{\bibfnamefont{P.}~\bibnamefont{Anderson}},
  \bibinfo{journal}{Physical Review} \textbf{\bibinfo{volume}{109}},
  \bibinfo{pages}{1492} (\bibinfo{year}{1958}).

\bibitem[{\citenamefont{Mahmoudian et~al.}(2014)\citenamefont{Mahmoudian, Tang,
  and Dobrosavljevi{\'c}}}]{mahmoudian2014quantum}
\bibinfo{author}{\bibfnamefont{S.}~\bibnamefont{Mahmoudian}},
  \bibinfo{author}{\bibfnamefont{S.}~\bibnamefont{Tang}}, \bibnamefont{and}
  \bibinfo{author}{\bibfnamefont{V.}~\bibnamefont{Dobrosavljevi{\'c}}},
  \bibinfo{journal}{arXiv preprint arXiv:1411.4558}  (\bibinfo{year}{2014}).

\bibitem[{\citenamefont{Dobrosavljevi{\'c} and
  Kotliar}(1997)}]{dobrosavljevic1997mean}
\bibinfo{author}{\bibfnamefont{V.}~\bibnamefont{Dobrosavljevi{\'c}}}
  \bibnamefont{and} \bibinfo{author}{\bibfnamefont{G.}~\bibnamefont{Kotliar}},
  \bibinfo{journal}{Physical review letters} \textbf{\bibinfo{volume}{78}},
  \bibinfo{pages}{3943} (\bibinfo{year}{1997}).

\bibitem[{\citenamefont{Aguiar et~al.}(2009)\citenamefont{Aguiar,
  Dobrosavljevi\ifmmode~\acute{c}\else \'{c}\fi{}, Abrahams, and
  Kotliar}}]{PhysRevLett.102.156402}
\bibinfo{author}{\bibfnamefont{M.~C.~O.} \bibnamefont{Aguiar}},
  \bibinfo{author}{\bibfnamefont{V.}~\bibnamefont{Dobrosavljevi\ifmmode~\acute{c}\else
  \'{c}\fi{}}}, \bibinfo{author}{\bibfnamefont{E.}~\bibnamefont{Abrahams}},
  \bibnamefont{and} \bibinfo{author}{\bibfnamefont{G.}~\bibnamefont{Kotliar}},
  \bibinfo{journal}{Phys. Rev. Lett.} \textbf{\bibinfo{volume}{102}},
  \bibinfo{pages}{156402} (\bibinfo{year}{2009}).

\bibitem[{\citenamefont{Aguiar et~al.}(2003)\citenamefont{Aguiar, Miranda, and
  Dobrosavljevi\'{c}}}]{aguiaretal1}
\bibinfo{author}{\bibfnamefont{M.~C.~O.} \bibnamefont{Aguiar}},
  \bibinfo{author}{\bibfnamefont{E.}~\bibnamefont{Miranda}}, \bibnamefont{and}
  \bibinfo{author}{\bibfnamefont{V.}~\bibnamefont{Dobrosavljevi\'{c}}},
  \bibinfo{journal}{Phys. Rev. B} \textbf{\bibinfo{volume}{68}},
  \bibinfo{pages}{125104} (\bibinfo{year}{2003}).

\bibitem[{\citenamefont{Goldenfeld}(1992)}]{goldenfeld1992lectures}
\bibinfo{author}{\bibfnamefont{N.}~\bibnamefont{Goldenfeld}}
  (\bibinfo{year}{1992}).

\bibitem[{\citenamefont{Efros and Shklovskii}(1975)}]{efros1975coulomb}
\bibinfo{author}{\bibfnamefont{A.}~\bibnamefont{Efros}} \bibnamefont{and}
  \bibinfo{author}{\bibfnamefont{B.}~\bibnamefont{Shklovskii}},
  \bibinfo{journal}{Journal of Physics C: Solid State Physics}
  \textbf{\bibinfo{volume}{8}}, \bibinfo{pages}{L49} (\bibinfo{year}{1975}).

\bibitem[{\citenamefont{Efros}(1976)}]{efros1976coulomb}
\bibinfo{author}{\bibfnamefont{A.}~\bibnamefont{Efros}},
  \bibinfo{journal}{Journal of Physics C: Solid State Physics}
  \textbf{\bibinfo{volume}{9}}, \bibinfo{pages}{2021} (\bibinfo{year}{1976}).

\bibitem[{\citenamefont{Efros and Pollak}(1985)}]{efros1985electron}
\bibinfo{author}{\bibfnamefont{A.~L.} \bibnamefont{Efros}} \bibnamefont{and}
  \bibinfo{author}{\bibfnamefont{M.}~\bibnamefont{Pollak}},
  \emph{\bibinfo{title}{Electron-electron interactions in disordered systems}}
  (\bibinfo{publisher}{North Holland}, \bibinfo{year}{1985}).

\bibitem[{\citenamefont{M.Janssen}(1998)}]{re:Janssen98}
\bibinfo{author}{\bibnamefont{M.Janssen}}, \bibinfo{journal}{Phys. Rep.}
  \textbf{\bibinfo{volume}{295}}, \bibinfo{pages}{1} (\bibinfo{year}{1998}).

\bibitem[{\citenamefont{Amini et~al.}(2013)\citenamefont{Amini, Kravtsov, and
  Mueller}}]{amini2013multifractality}
\bibinfo{author}{\bibfnamefont{M.}~\bibnamefont{Amini}},
  \bibinfo{author}{\bibfnamefont{V.}~\bibnamefont{Kravtsov}}, \bibnamefont{and}
  \bibinfo{author}{\bibfnamefont{M.}~\bibnamefont{Mueller}},
  \bibinfo{journal}{arXiv preprint arXiv:1305.0242}  (\bibinfo{year}{2013}).

\bibitem[{\citenamefont{Dobrosavljevic
  et~al.}(2003)\citenamefont{Dobrosavljevic, Pastor, and
  Nikolic}}]{0295-5075-62-1-076}
\bibinfo{author}{\bibfnamefont{V.}~\bibnamefont{Dobrosavljevic}},
  \bibinfo{author}{\bibfnamefont{A.~A.} \bibnamefont{Pastor}},
  \bibnamefont{and} \bibinfo{author}{\bibfnamefont{B.~K.}
  \bibnamefont{Nikolic}}, \bibinfo{journal}{EPL (Europhysics Letters)}
  \textbf{\bibinfo{volume}{62}}, \bibinfo{pages}{76} (\bibinfo{year}{2003}).

\bibitem[{\citenamefont{Terletska et~al.}(2014)\citenamefont{Terletska, Ekuma,
  Moore, Tam, Moreno, and Jarrell}}]{PhysRevB.90.094208}
\bibinfo{author}{\bibfnamefont{H.}~\bibnamefont{Terletska}},
  \bibinfo{author}{\bibfnamefont{C.~E.} \bibnamefont{Ekuma}},
  \bibinfo{author}{\bibfnamefont{C.}~\bibnamefont{Moore}},
  \bibinfo{author}{\bibfnamefont{K.-M.} \bibnamefont{Tam}},
  \bibinfo{author}{\bibfnamefont{J.}~\bibnamefont{Moreno}}, \bibnamefont{and}
  \bibinfo{author}{\bibfnamefont{M.}~\bibnamefont{Jarrell}},
  \bibinfo{journal}{Phys. Rev. B} \textbf{\bibinfo{volume}{90}},
  \bibinfo{pages}{094208} (\bibinfo{year}{2014}).

\bibitem[{\citenamefont{Ekuma et~al.}(2014)\citenamefont{Ekuma, Terletska, Tam,
  Meng, Moreno, and Jarrell}}]{PhysRevB.89.081107}
\bibinfo{author}{\bibfnamefont{C.~E.} \bibnamefont{Ekuma}},
  \bibinfo{author}{\bibfnamefont{H.}~\bibnamefont{Terletska}},
  \bibinfo{author}{\bibfnamefont{K.-M.} \bibnamefont{Tam}},
  \bibinfo{author}{\bibfnamefont{Z.-Y.} \bibnamefont{Meng}},
  \bibinfo{author}{\bibfnamefont{J.}~\bibnamefont{Moreno}}, \bibnamefont{and}
  \bibinfo{author}{\bibfnamefont{M.}~\bibnamefont{Jarrell}},
  \bibinfo{journal}{Phys. Rev. B} \textbf{\bibinfo{volume}{89}},
  \bibinfo{pages}{081107} (\bibinfo{year}{2014}).

\bibitem[{\citenamefont{Georges et~al.}(1996)\citenamefont{Georges, Kotliar,
  Krauth, and Rozenberg}}]{georges1996dynamical}
\bibinfo{author}{\bibfnamefont{A.}~\bibnamefont{Georges}},
  \bibinfo{author}{\bibfnamefont{G.}~\bibnamefont{Kotliar}},
  \bibinfo{author}{\bibfnamefont{W.}~\bibnamefont{Krauth}}, \bibnamefont{and}
  \bibinfo{author}{\bibfnamefont{M.}~\bibnamefont{Rozenberg}},
  \bibinfo{journal}{Reviews of Modern Physics} \textbf{\bibinfo{volume}{68}},
  \bibinfo{pages}{13} (\bibinfo{year}{1996}).

\bibitem[{\citenamefont{Dobrosavljevi{\'c}
  et~al.}(2003)\citenamefont{Dobrosavljevi{\'c}, Pastor, and
  Nikolic}}]{pastor2001tmt}
\bibinfo{author}{\bibfnamefont{V.}~\bibnamefont{Dobrosavljevi{\'c}}},
  \bibinfo{author}{\bibfnamefont{A.~A.} \bibnamefont{Pastor}},
  \bibnamefont{and} \bibinfo{author}{\bibfnamefont{B.~K.}
  \bibnamefont{Nikolic}}, \bibinfo{journal}{Europh. Lett.}
  \textbf{\bibinfo{volume}{62}}, \bibinfo{pages}{76} (\bibinfo{year}{2003}).

\bibitem[{\citenamefont{Dobrosavljevic}(2010)}]{dobrosavljevic2010typical}
\bibinfo{author}{\bibfnamefont{V.}~\bibnamefont{Dobrosavljevic}},
  \bibinfo{journal}{50 Years of Anderson Localization} p. \bibinfo{pages}{425}
  (\bibinfo{year}{2010}).

\end{thebibliography}

\end{document}